\UseRawInputEncoding
\documentclass[%
 reprint,
 amsmath,amssymb,
 aps,
]{revtex4-2}

\usepackage{graphicx}
\usepackage{dcolumn}
\usepackage{bm}
\usepackage{caption}
\usepackage{subcaption}

\begin{document}

\preprint{APS/123-QED}

\title{Building robust surrogate models of laser-plasma interactions\\ using large scale PIC simulation}

\author{Nathan Smith}
 \email{Nathan.Smith@york.ac.uk}
\author{Kate Lancaster}
\author{Chris Ridgers}%
\author{Chris Arran}
\affiliation{%
 York Plasma Institute,\\
 University of York
}%

\author{Stuart Morris}
\affiliation{University of Warwick}





\begin{abstract}
As the repetition rates of ultra-high intensity lasers increase, simulations used for the prediction of experimental results may need to be augmented with machine learning to keep up. In this paper, the usage of gaussian process regression in producing surrogate models of laser-plasma interactions from particle-in-cell simulations is investigated. Such a model retains the characteristic behaviour of the simulations but allows for faster on-demand results and estimation of statistical noise. A demonstrative model of Bremsstrahlung emission by hot electrons from a femtosecond timescale laser pulse in the $10^{20} - 10^{23}\;\mathrm{Wcm}^{-2}$ intensity range is produced using 800 simulations of such a laser-solid interaction from 1D hybrid-PIC. While the simulations required 84,000 CPU-hours to generate, subsequent training occurs on the order of a minute on a single core and prediction takes only a fraction of a second. The model trained on this data is then compared against analytical expectations. The efficiency of training the model and its subsequent ability to distinguish types of noise within the data are analysed, and as a result error bounds on the model are defined. 
\end{abstract}

\maketitle


\section{\label{sec:intro}Introduction \& Background}

With next-generation high-intensity, high-repetition rate laser facilities set to begin operation soon, multiple new avenues for research in high intensity laser-matter interactions and available secondary sources are beginning to open\cite{gales2015laser}. High repetition rates allow for vastly larger scale explorations of the available parameter spaces\cite{Furch_2022}. With both high intensity and high repetition rates, large parameter scans can be done to study unexplored regimes such as strong field QED\cite{baumann2019laser, PhysRevLett.127.114801, Macleod_2022}, sophisticated manufacturing methods can be used in industry\cite{microfabrication, laser_ablation}, and production of secondary sources allows for imaging of other systems\cite{hadrich2015exploring}.  Alongside this, it would be beneficial to have quick, efficient models to both predict the outcomes of experiments using these lasers and find optimum sets of parameters for them. 

The standard method of modelling high-intensity laser-plasma interactions is through Particle-In-Cell (PIC) and hybrid-PIC simulations, however these are typically too slow for in-situ modelling and have inherent variability due to statistical noise. The latter can be improved through convergence testing, however this further increases the computational cost of performing simulations. Because of this, we are interested in developing a surrogate model, where simulations are done across the relevant parameter space up-front, from which a model is created that can be used to estimate the result of a PIC simulation, as well as an associated uncertainty on the result, at any set of parameters within that space. Important to note is that this is an interpolative system, and as such is limited to the region simulated, quickly giving results with large error bars as the input moves outside the region. Surrogate models such as this can be made with many different approaches, and are used in the many fields that require extensive use of computationally-intense simulation, namely engineering\cite{alizadeh2020managing}. In the case of laser-plasma interactions, a surrogate model should (1) be far quicker than an equivalent PIC simulation of the system, (2) reliably interpolate a sparse dataset to produce accurate models and (3) produce reliable estimates of the error, both due to the statistical nature of the underlying simulations and due to the sparse sampling of the parameter space.

Previous work in this field has looked at using such methods for optimization and control of experiments, for instance in maximising the generation of synchrotron radiation in laser-solid interactions \cite{Goodman_King_Dolier_Wilson_Gray_McKenna_2023, Goodman_King_Dolier_Wilson_Gray_McKenna_2024}, modelling the properties of electron bunches in laser wakefield acceleration \cite{Streeter_Colgan_Cobo_Arran_Los_Watt_Bourgeois_Calvin_Carderelli_Cavanagh_et_al._2023} and calculating emission probabilities for the Breit-Wheeler process \cite{PhysRevAccelBeams.26.054601}. Our work focuses on generalising this approach using gaussian processes, and from it developing a methodology to use in the future for approximate modelling of laser-plasma systems. We believe this methodology, especially following planned further development, is an elegant solution to the problems posed above, notably in the treatment of noise and uncertainty in the system.

In this paper we produce a surrogate model based on the results of large scale hybrid-PIC simulations of a laser-solid interaction, and evaluate this based on its ability to reproduce known results, the quantification of statistical uncertainty across the parameter space, and how these qualities vary with hyperparameters of the underlying simulations, namely resolution. To test the efficacy of the method, we looked at the physics of bremsstrahlung emission in a plastic target. This was chosen due to this setup having simple scaling laws for the total bremsstrahlung radiation generated allowing for easy comparison, as well as the comparative ease of simulating the system. 

The rest of this section, (\ref{Bremss}), covers the physics behind bremsstrahlung emission and introduces some simple scaling laws which we expect our model to be able to reproduce. In Section \ref{sec:method} we cover specifics of the system being looked at and how it is simulated, followed by how the resulting surrogate model is created through a gaussian process regression method. In Section \ref{sec:results} we cover the overall features of the resultant dataset and the surrogate model produced from it. We also cover variations of this surrogate model resulting from changes to the simulation set-up, to analyze how robust this method to such changes and the statistical nature of PIC simulation. Finally in Section \ref{sec:discuss} we look at the utility and efficiency of this approach to modelling laser-plasma interactions, and on how future work could expand on this to make use of the scheme in experiments of a similar nature.

\subsection{\label{Bremss} Bremsstrahlung}
Bremsstrahlung is the radiation produced by charged particles interacting with the electric fields of nuclei, and a full semi-classical description of the process can be found in \cite{JacksonJohnDavid1925-20161999Ce/J}. In our case, bremsstrahlung comes in the form of x-rays emitted by accelerated electrons as they travel through and are decelerated by the atoms and ions in the target material. A fraction of the electrons at the front surface of the target will be accelerated by the laser pulse to much higher energies than the rest of the electron population and are known as ``hot" electrons. Each of these electrons, with mass $m_e$, have an energy $\epsilon$ that approximately follows an Boltzmann distribution with a mean given by 
\begin{equation}\label{mean_energy}
    \langle \epsilon \rangle = a_0m_ec^2,\quad a_0 = \frac{eE_0}{m_ec\omega_L},
\end{equation}
where $a_0$ is the normalized laser amplitude, $E_0$ is the peak electric field strength and $\omega_L$ is the angular frequency of the laser. Assuming a constant laser-to-electron conversion efficiency $\eta_{l\rightarrow e}$(this parameter generally varies slowly with the system parameters as well as surface geometry, typically reaching values on the order of $10-30\%$\cite{PhysRevLett.105.235001}, but is kept constant at $\eta_{l\rightarrow e}=0.3$ in our simulations), this means that the total number of hot electrons injected by a constant laser intensity $I_0$ incident on an area $\delta A$ for a period $\tau$ is given by
\begin{equation}
    N_e = \frac{I_0\tau\eta_{l\rightarrow e}\delta A}{\langle\epsilon\rangle}.
\end{equation}

At the laser intensities looked at in this study  $a_0\gg 1$ and so our hot electron population is relativistic. In such a situation, the differential radiation cross section for an electron travelling within a material with atomic number $Z$ is given in CGS units by
\begin{equation}\label{cross_sec_screened}
    \frac{d\chi}{d\omega} = \frac{16Z^2e^6}{3m_e^2c^5}\log\left(\frac{233}{Z^{1/3}}\right),
\end{equation}
where $\omega$ is the frequency of the emitted photon and $\chi(\omega)$ is the sum of all possible cross sections for emitting a photon of frequency up to $\omega$, multiplied by their respective radiated energy.
In Equation (\ref{cross_sec_screened}) we have assumed the screening of the electric field of the nuclei by their surrounding electrons to be complete, as in \cite{JacksonJohnDavid1925-20161999Ce/J}, and thus the cross section becomes constant with the frequency, which occurs for very high electron energies. It should be noted that this only holds for frequencies up to a certain value, $\omega_{max}$, which approximately corresponds to the total energy of the electron, $\hbar\omega_{max}\simeq\epsilon=\gamma m_ec^2$.

We can integrate this cross section for all frequencies up to the maximum frequency and multiply the material number density $n_i$ to find the energy radiated per unit distance travelled by the electron as
\begin{equation}
    \frac{dE_{brems}}{dx} = n_i\int^{\omega_{max}}_0 \frac{d\chi}{d\omega} d\omega.
\end{equation}
In the case of both screened and unscreened cross sections, the result scales with the total energy of the electron. Finally, we can see that this emission will continue for as long as the electron travels through the material, and so we expect more bremsstrahlung emission for large target depths. From this we can infer some simple scaling laws we expect any model we produce to follow.
\begin{equation}
\eta \propto n_i{I_0}^{1/2}d 
\end{equation}
This simple model can be used to benchmark our final model to see if it matches expectations. Note that this does not take in to account other energy loss effects and simplifies some elements of the physics (such as the assumed total screening), and so the the actual scaling of our model with these parameters will not be exact. Additionally, equating the electron path length with target depth is not entirely accurate, as refluxing at the target edges will increase the electron path length, albeit with some energy loss at each scattering.

\section{\label{sec:method}Methodology}
Our study was performed in two stages; first, 800 simulations of a laser solid interaction were performed using the 1D version of Hybrid-EPOCH, a PIC code where the material and its cold electrons are treated as a background field  rather than as a set of ion and electron macro-particles. This significantly speeds up computation times by allowing a lower resolution to be used and only needing to track the hot electron population. This code is described further in \cite{10.1063/5.0055398}. This set was repeated with differing grid resolution. Secondly, these datasets were used with a Gaussian Process regression framework to create surrogate models, which are then evaluated based on the bremsstrahlung physics laid out in (\ref{Bremss}). 
 
\subsection{\label{subsec:sims}Simulation set-up} 
Our simulation setup consists of a 1D space filled with plastic, modelled by uniformly distributed Carbon and Hydrogen in a $1:2$ ratio, respectively. The size of this space is the depth of the target $d$, while the densities of the two material components sum to the total number density $n_i$. The incident laser is not directly simulated, instead electrons are injected into the simulation space directly with properties that are based on the given laser parameters. These electrons are initialized with energies following a Boltzmann distribution such that the probability density function of the electron energies, $p(\epsilon)$, is given by 
\begin{equation}
    p(\epsilon) = \frac{1}{\langle\epsilon\rangle}\exp\left(-\frac{\epsilon}{\langle\epsilon\rangle}\right),
\end{equation}
where the mean electron energy $\langle\epsilon\rangle$ is equal to that given in Equation (\ref{mean_energy}). The definition of the normalized field amplitude can be rewritten as $a_0 = 8.5\times 10^{-6}\sqrt{I_0\lambda^2}$, where $I_0$ is the laser intensity, and $\lambda$ is the laser wavelength. The incoming laser has a 800nm wavelength. The number of electrons injected per time step matches the laser's Gaussian-shaped temporal intensity profile, with starting and end cut-offs two standard deviations in both cases from the intensity peak. This standard deviation (i.e. the pulse timescale) is set as the random parameter $\tau$, and so the laser time profile has a FWHM of $2.355\tau$. Due to the 1D nature of the simulations, no spatial profile is used for the laser/electron injection, with simulation properties assumed to be uniform in the unsimulated transverse dimensions by the code. Macroparticles are therefore weighted assuming that each 1D cell represents a 3D cell with transverse extents of 1m.

In a typical laser plasma interaction, the hot electrons are ejected out of the rear of the target in greater numbers than ions. This results in a charge imbalance and thus creates an electric field at the rear of the target (called the sheath field) that accelerates ions and reflects many of the hot electrons back in to the target with some energy loss and scattering \cite{Rusby_Armstrong_Scott_King_McKenna_Neely_2019}. As this requires ions to be simulated, this effect can not be directly simulated in hybrid-EPOCH, so instead its effects are approximated by use of TNSA boundary conditions. These have three defined parameters; an escape energy, scattering energy loss, and scattering angle. The electric field strength, and as such the energy required to escape the target, is proportional to the hot electron temperature, and so the escape energy parameter is calculated as proportional to the mean electron energy, up to a constant $\kappa_{esc}$, which is defined as $1.5$ in these simulations. The scattering energy loss is calculated in the same manner, but with a proportionality constant of $\kappa_{TNSA}=0.0027$. The scattering angle is kept at a constant $20^o$. These choices are based on previously done standard EPOCH simulations \cite{10.1063/5.0055398} and are discussed further later. 

\begin{table}[h]
    \centering
    \begin{tabular}{|c||c|c|c|}
        \hline
        Parameter & Minimum & Maximum & Units  \\
        \hline
        Intensity & $10^{20}$ & $10^{23}$ & Wcm$^{-2}$\\
        Pulse Length & 1 & 1000 & fs \\
        Target Depth & 1 & 100 & $\mu$m \\
        Number Density & $10^{28}$ & $10^{30}$ & m$^{-3}$\\
        \hline
    \end{tabular}
    \caption{Variable simulation parameters, the limits used for them and the units of each.}
    \label{tab:parameters}
\end{table}

There are four parameters of interest that are varied from simulation to simulation. These are listed above in Table \ref{tab:parameters}. Each parameter is sampled from a log-uniform distribution between their listed maxima and minima. The hot electrons in each simulation emit high energy photons, the dynamics of which are frozen so that they do not leave the simulation. Additionally, while the simulation is only 1D, the photons and electrons have momentum calculated for all three spatial dimensions. The electron dynamics and Bremsstrahlung emission are simulated for $0.5\ \textrm{ns}$ in all simulations.  

Additionally, at each point sampled in parameter space the simulations were performed twice, once with 40nm grid spacing and once with 100nm spacing, in order to investigate the effect on the resulting surrogate model when using simulations with lower resolution and as such increased noise. The timestep used in these simulations is calculated as the time taken for light to traverse a single cell multiplied by $0.8$, and is kept constant throughout the simulation. As such the 40nm and 100nm simulations have timesteps of approximately 0.107fs and 0.267fs, respectively. 

\subsection{\label{subsec:GPR}Modelling the results}
The collected simulation data is used to fit a model via Gaussian Process Regression (GPR).
Gaussian processes are a generalisation of the gaussian distribution, where functions themselves are sampled. This can be viewed as the limit of a multivariate gaussian distribution as the number of variables is taken to infinity, where each variable reflects the value of the function at some point. This effectively produces a probability distribution of possible functions. In this generalisation the covariance matrix of the distribution becomes what is known as a kernel function, $k(\mathbf{x}_i, \mathbf{x}_j)$, relating how correlated the value of a sampled function at a point $\mathbf{x}_i$, is to that at another point $\mathbf{x}_j$. The choice of this function will determine the form of any function sampled from the resultant gaussian process. When datapoints are observed, a conditional distribution is calculated from the overall distribution (called the prior distribution in the manner of Bayesian inference) using the observed values, giving a new gaussian process (i.e. the posterior) that accounts for the new observation. Furthermore, the hyperparameters of the kernel function are optimized using this observed data, varying them such that the log marginal likelihood of observing the set of datapoints is maximised.


\begin{figure*}[t!]
    \centering
    \begin{subfigure}{0.45\textwidth}
        \centering
        \includegraphics[width=\textwidth]{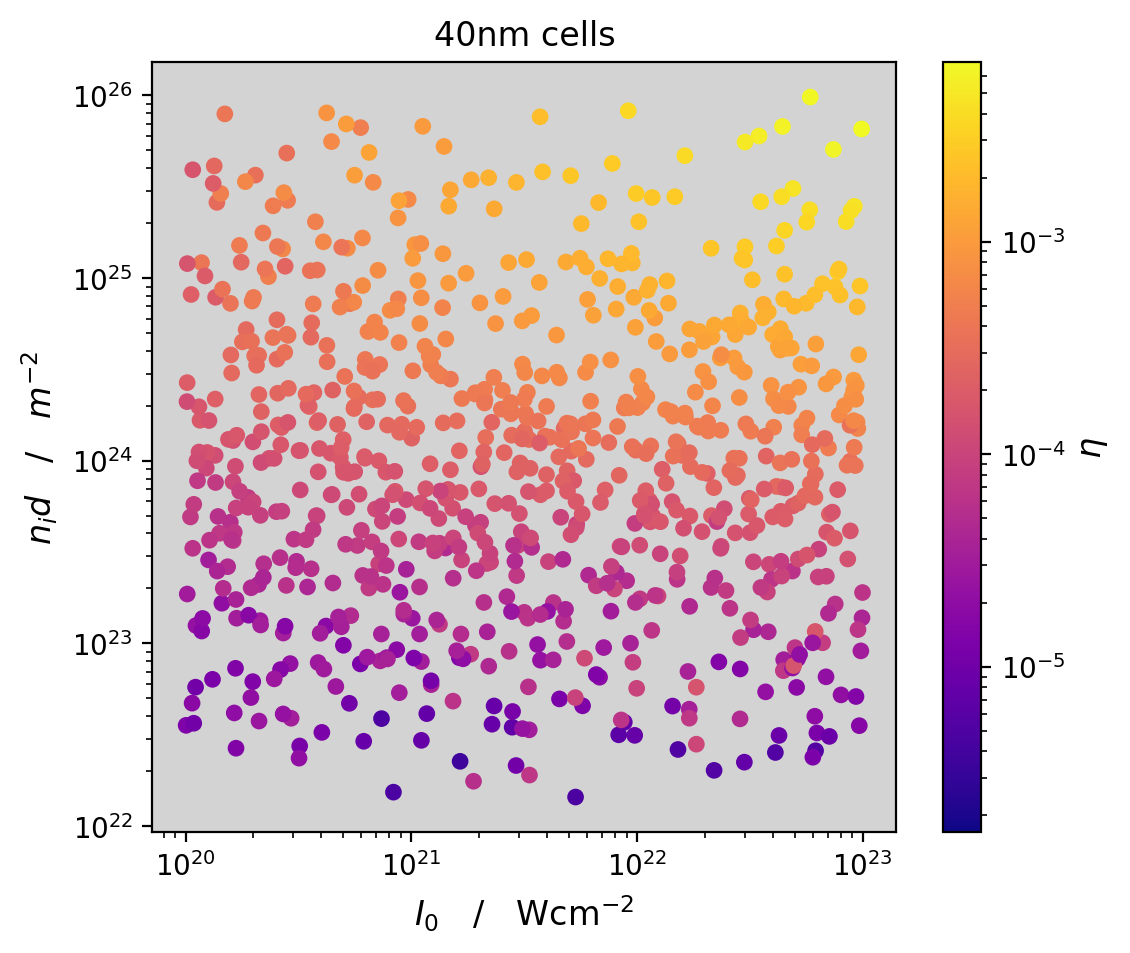}
        \caption{Summary of dataset with 40nm grid cells}
        \label{fig:summary_1}
    \end{subfigure}
    \hfill
    \begin{subfigure}{0.45\textwidth}
        \centering
        \includegraphics[width=\textwidth]{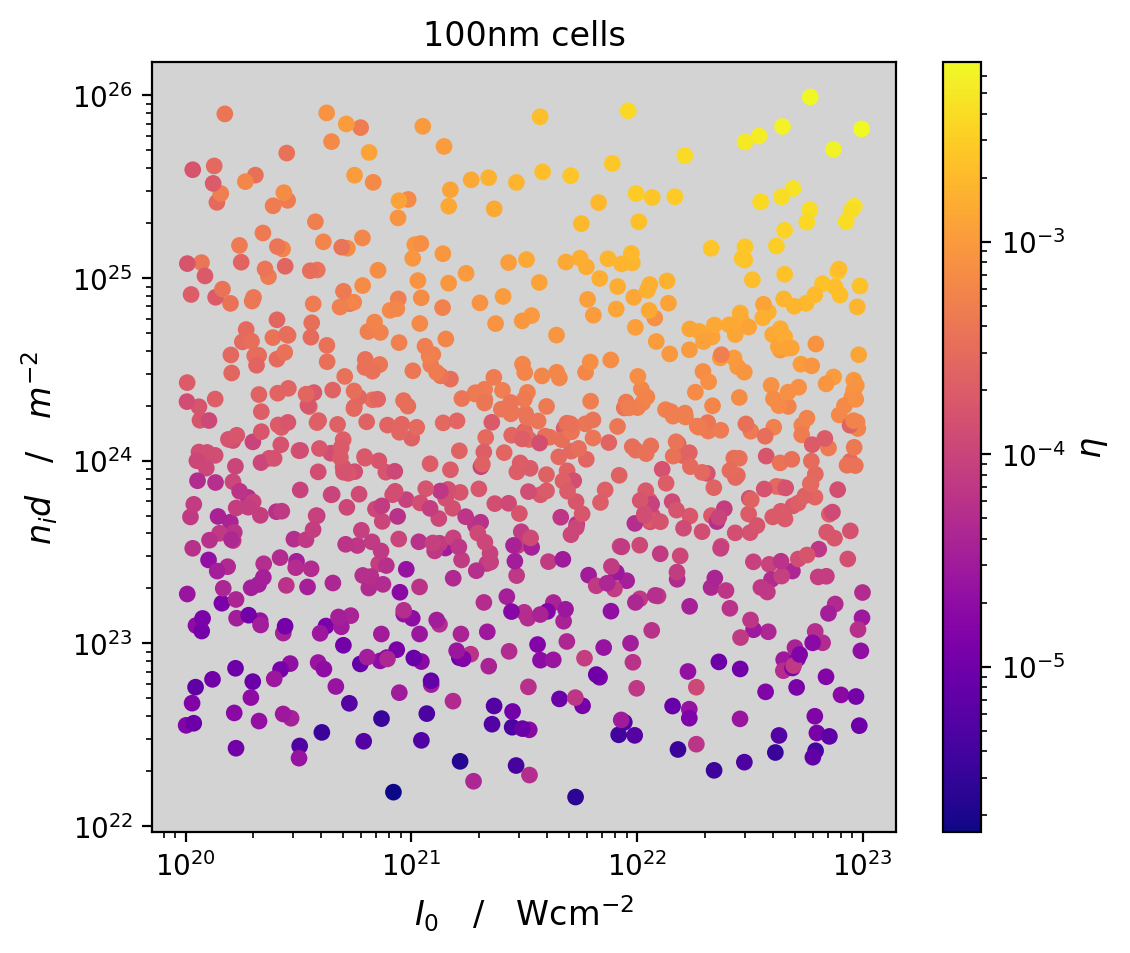}
        \caption{Summary of dataset with 100nm grid cells.}
        \label{fig:summary_2}
    \end{subfigure}
    \caption{Plots of the conversion efficiencies calculated at randomly sampled points across the simulation parameter space, for the same parameters in simulation with (a) 40nm grid cells and (b) 100nm grid cells. Pulse length is omitted from the plot axes as the conversion efficiency was found to not vary with this. Peak laser intensity is across the x-axis while the product of the target depth and target density is across the y-axis. }
    \label{fig:summaries}
\end{figure*}

In this study, gaussian processes were used to model the variation of laser-to-bremsstrahlung conversion efficiency across the parameter space for a simulated laser-solid interaction. We expect this to be a slowly varying function where values are locally correlated but with some amount of noise present. Because of these requirements our model made use of a square exponential function kernel with an added white noise kernel to account for the noise present in the simulations. The resulting kernel function between two points in parameter space $\mathbf{x_i}, \mathbf{x_j}$ is given by
\begin{equation}
    k(\mathbf{x_i}, \mathbf{x_j}) = a^2\exp\left(-\frac{|\mathbf{x_i} - \mathbf{x_j}|^2}{2l^2}\right)+w\delta(\mathbf{x_i},\mathbf{x_j}),
\end{equation}
where $\delta(a, b)$ is a function that equals unity when $a=b$ and zero otherwise, and $a, l$ and $w$ are the kernel hyperparameters, representing function magnitude, length scale and noise respectively. These hyperparameters are those optimized during fitting to find the Gaussian Process which produces functions that best match the data. In addition to the kernel hyperparameters, an error on the training data can be defined with the parameter $\alpha$. This acts similarly to the white kernel described above but is added solely for the correlation at training points, and is used in cases where target data has some uncertainty but the underlying function is not expected to be noisy. 

The conversion efficiency, i.e. the total energy of bremsstrahlung photons produced divided by the total input energy, is used as the output of the model. The model itself is logarithmic in nature, taking the logarithm of the input parameters and giving the total energy and relevant uncertainty in logarithmic form. This is due to the widely varying values of the parameters and resultant amount of bremsstrahlung emission. The assumed uncertainty, $\alpha$, on data points for fitting (as the true uncertainty is something we wish the model to produce) is tuned to prevent over fitting of the gaussian process to the data.  Below a critical value there is a trade off between uncertainty in the data points and that from the white kernel, while above the critical value the model overestimates the error in the simulations. Following training of the model, expectation values and the variance of our parameter of interest can be found at any point within the parameter space. Further information on the exact details of gaussian processes can be found in \cite{RasmussenCarlEdward2006Gpfm}.

Separate models are fitted for differing numbers of grid cells, differing numbers of data points and across many values of $\alpha$ in order to investigate the efficacy of the method. 

\section{\label{sec:results}Results}
The Simulations were run over multiple parallel jobs on the Viking2 cluster. The number of processing cores used by each simulation varied based on the simulation parameters. In total the 40nm set required 84,000 CPU-hours whilst the 100nm set took 14,000-CPU hours.
Once all simulations were completed, the total energy emitted as bremsstrahlung is calculated from the sum of photon energies multiplied by the weights of their respective particles. The total Laser energy is calculated assuming a uniform $1\mathrm{m} \times 1\mathrm{m}$ spatial distribution for the laser pulse for consistency with how macroparticles are weighted by the 1D code. The ratio between these two values is then calculated as the Laser to Bremsstrahlung conversion efficiency. Figure \ref{fig:summary_1} shows how this conversion efficiency varies across a reduced form of the parameter space.

\subsection{\label{subsec:GPR}Gaussian Process model}
\begin{figure*}
    \centering
    \includegraphics[width=\textwidth]{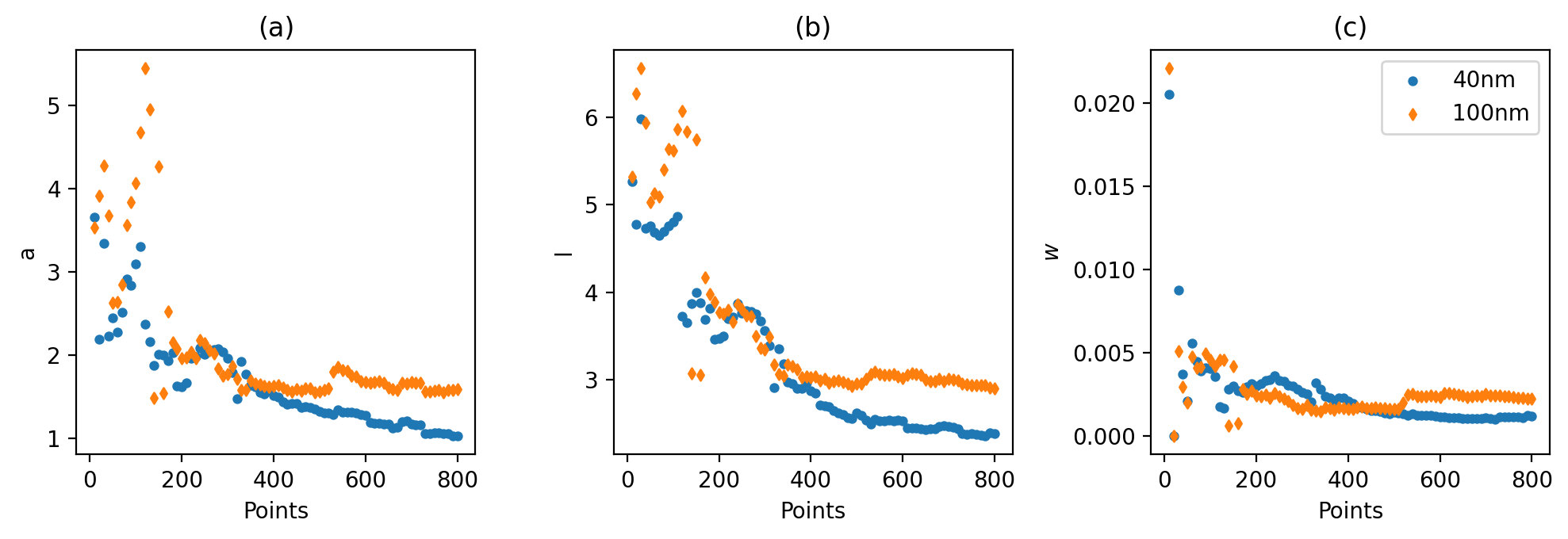}
    \caption{Convergence of the three kernel hyperparameters $a, l$ and $w$ in plots (a), (b) and (c), respectively.}
    \label{fig:param_convergence}
\end{figure*}

The dataset summarised in Figure \ref{fig:summary_1} was used to fit a GPR model. With a point uncertainty of $\alpha = 10^{-10}$ (effectively certain conversion efficiencies) and a training set of 800 simulation data points, the parameters of the optimized kernel function were found to be $a = 1.014, l = 2.377$ and $w=1.21\times10^{-3}$. This training process took 82 seconds, with subsequent prediction of 10,000 datapoints requiring 0.7 seconds, both on a single CPU core. The convergence of these parameters as the number of simulation data points increases is shown in Figure(\ref{fig:param_convergence}).

\begin{figure*}[t]
    \centering
    \includegraphics[width=0.5\textwidth]{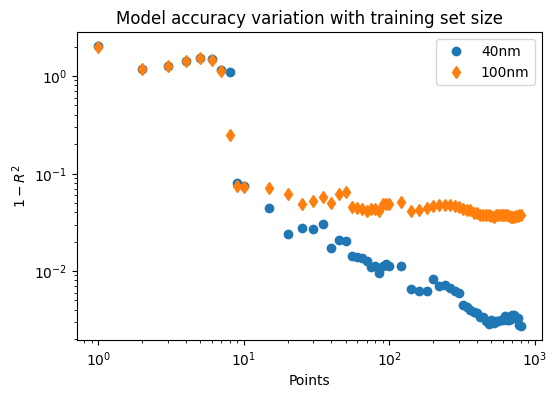}
    \caption{Variation of the model error, calculated as the difference between the coefficient of determination $R^2$ and its highest possible value, $1$. The results for models from both datasets are included and it can be seen that, following an initial drop in error at approximately 10 points, the error decreases following a scaling law.}
    \label{Accuracy}
\end{figure*}

Additionally, how the accuracy of the model improves in this scenario has been investigated using an additional set of 80 simulations. These are within the same parameter space as the training data, and have been generated in the same manner using a cell size of 40nm. The predictions of the Gaussian process at the parameter values of this test data are compared against the calculated simulation values using the coefficient of determination,
\begin{equation}
    R^2 = 1-\sum_i\frac{(\eta_i - \eta^{true}_i)^2}{\sigma_{true}^2},
\end{equation}
where $\eta_i$ and $\eta^{true}_i$ are the predicted and simulation calculated conversion efficiencies for set of parameters $i$, respectively, and $\sigma_{true}^2$ is the variance of the simulation calculated conversion efficiencies. A result of $R^2 = 1$ indicates a model that perfectly predicts the simulation results, while $R^2=0$ indicates one which exclusively returns the mean value of the true data, with $R^2 < 0$ in the case of a model performing worse than this. The results as the number of training samples increases, given in the form of the proximity to 1 as a measure of the model error, are given in Figure (\ref{Accuracy}). It should be noted that this coefficient of determination is calculated directly from the model outputs, i.e. in logarithmic space, and so the values calculated may not completely reflect those in real space. 

\begin{figure*}[t]
    \begin{subfigure}{0.49\textwidth}
        \centering
        \includegraphics[width=\textwidth]{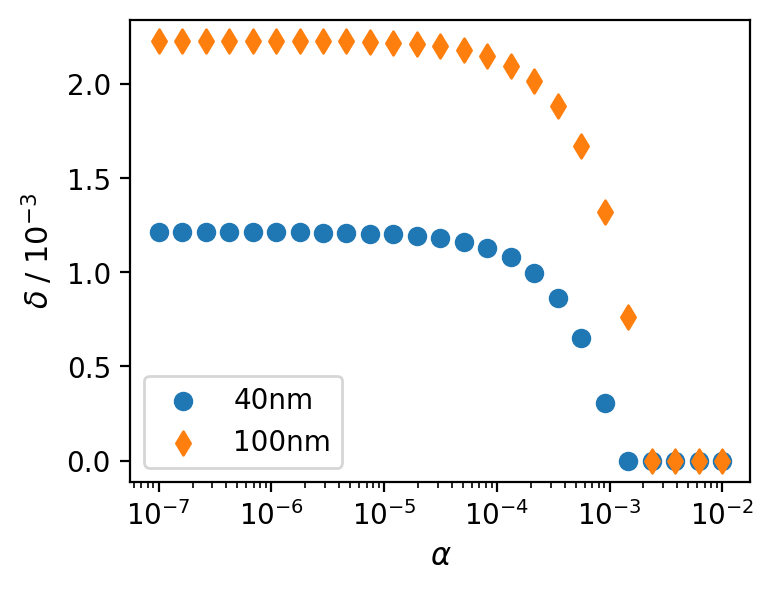}
        \caption{Logarithmic variation}
    \end{subfigure}
    \hfill
    \begin{subfigure}{0.49\textwidth}
        \centering
        \includegraphics[width=\textwidth]{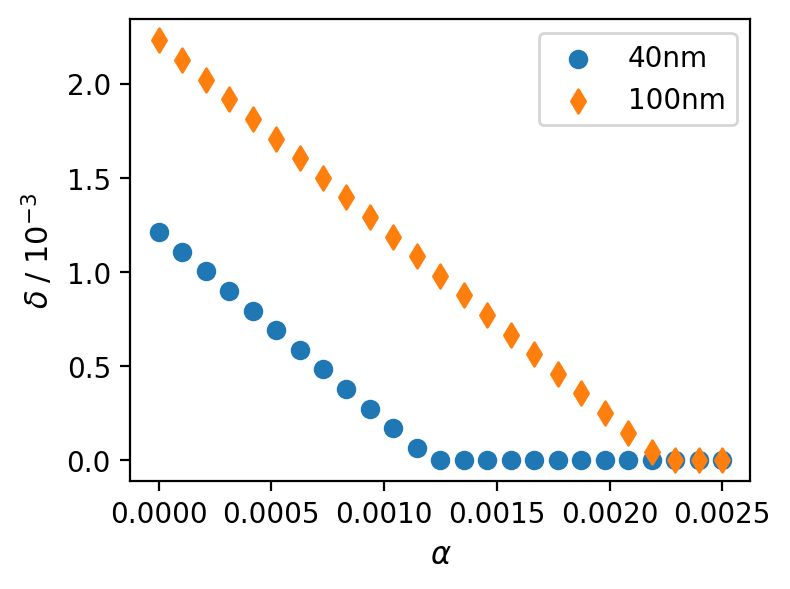}
        \caption{Variation for $\alpha \sim 10^{-3}$}
    \end{subfigure}
    \caption{Variation of the optimal value of the white noise hyperparameter $\delta$ as the target data uncertainty is varied. The variation over different orders of magnitude is shown in (a), while in (b), which shows the variation across a linear axis, it can be seen that $\alpha$ and $\delta$ sum to a constant value, the 'true' noise of the model. }
    \label{fig:alpha}
\end{figure*}

As the target uncertainty $\alpha$ is an undetermined parameter of the dataset and will have an impact on the calculated white noise of the model, a logarithmic scan of $\alpha$ is done to investigate how the resulting surrogate model changes. The optimum white noise kernel for each value of alpha is shown in Figure \ref{fig:alpha}. From this we can see that below a critical value, $\alpha$ and the noise present in the surrogate mode, $\delta$, will sum to this critical value, which can as such be considered to be the ``true" noise of the dataset used. Above this value, the model could be considered to be underfitting the data, however the mean prediction does not change significantly.

\begin{figure*}[t]
    \centering
    \begin{subfigure}{0.49\textwidth}
        \centering
        \includegraphics[width=\textwidth]{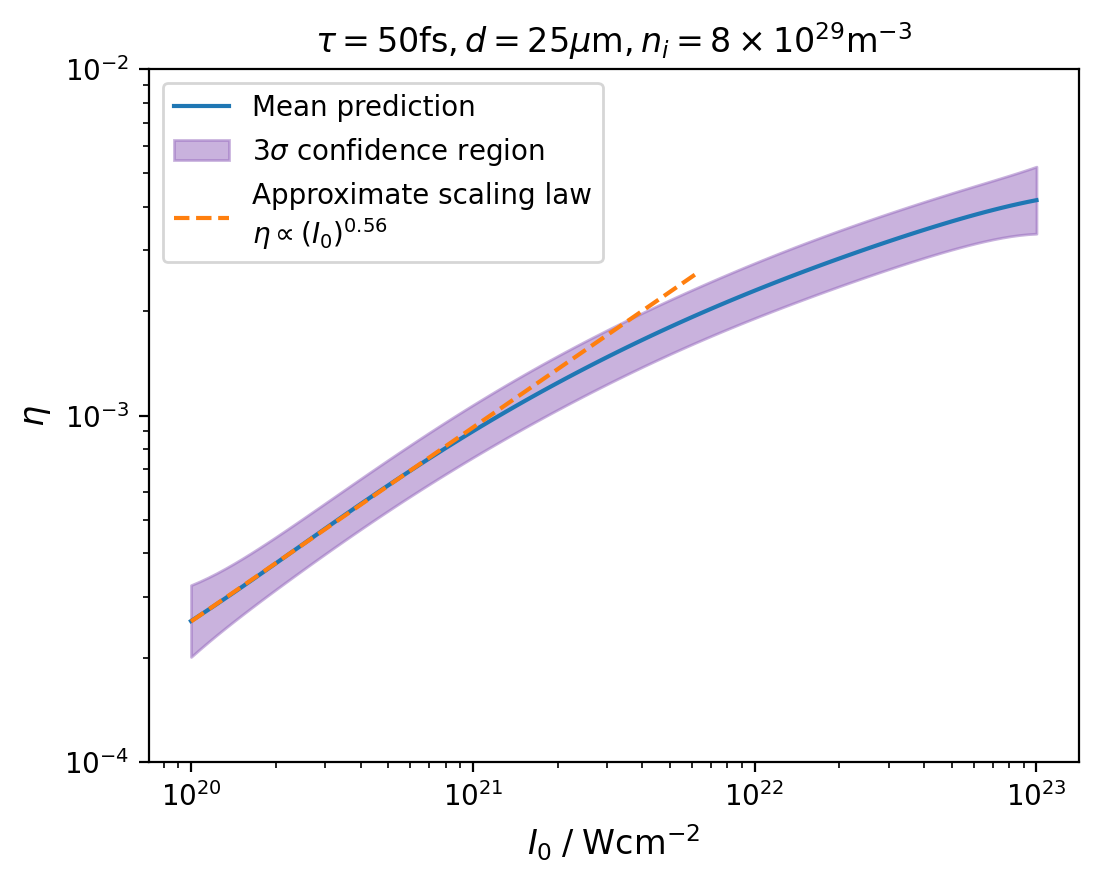}
        \caption{Intensity}
    \end{subfigure}
    \begin{subfigure}{0.49\textwidth}
        \centering
        \includegraphics[width=\textwidth]{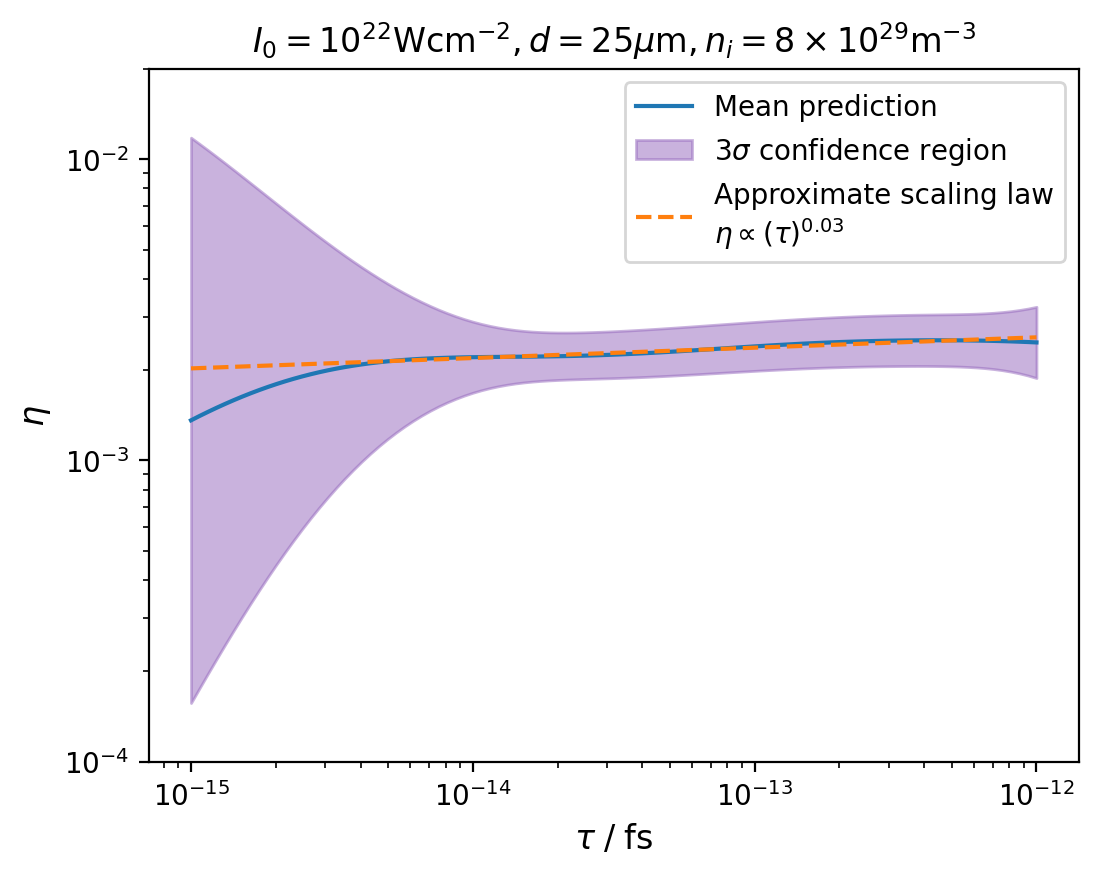}
        \caption{Pulse duration}
    \end{subfigure}
    \begin{subfigure}{0.49\textwidth}
        \centering
        \includegraphics[width=\textwidth]{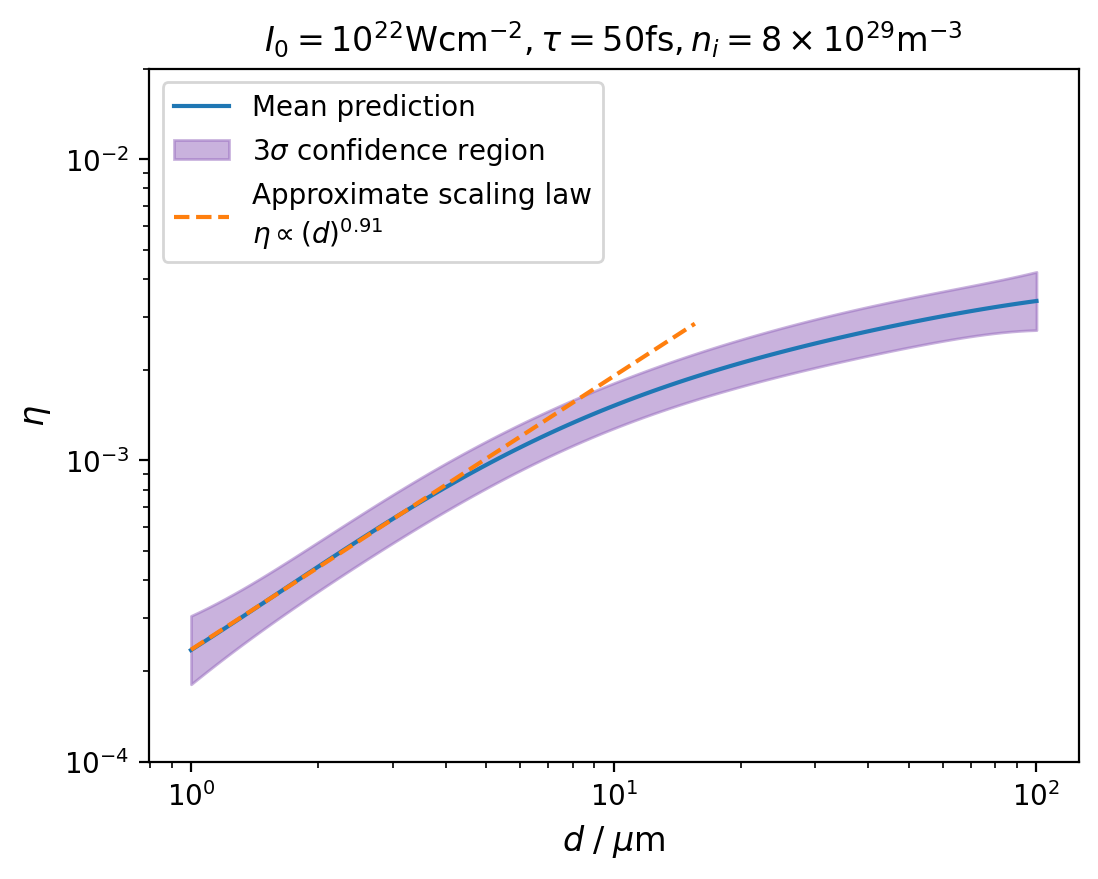}
        \caption{Target depth}
    \end{subfigure}
    \begin{subfigure}{0.49\textwidth}
        \centering
        \includegraphics[width=\textwidth]{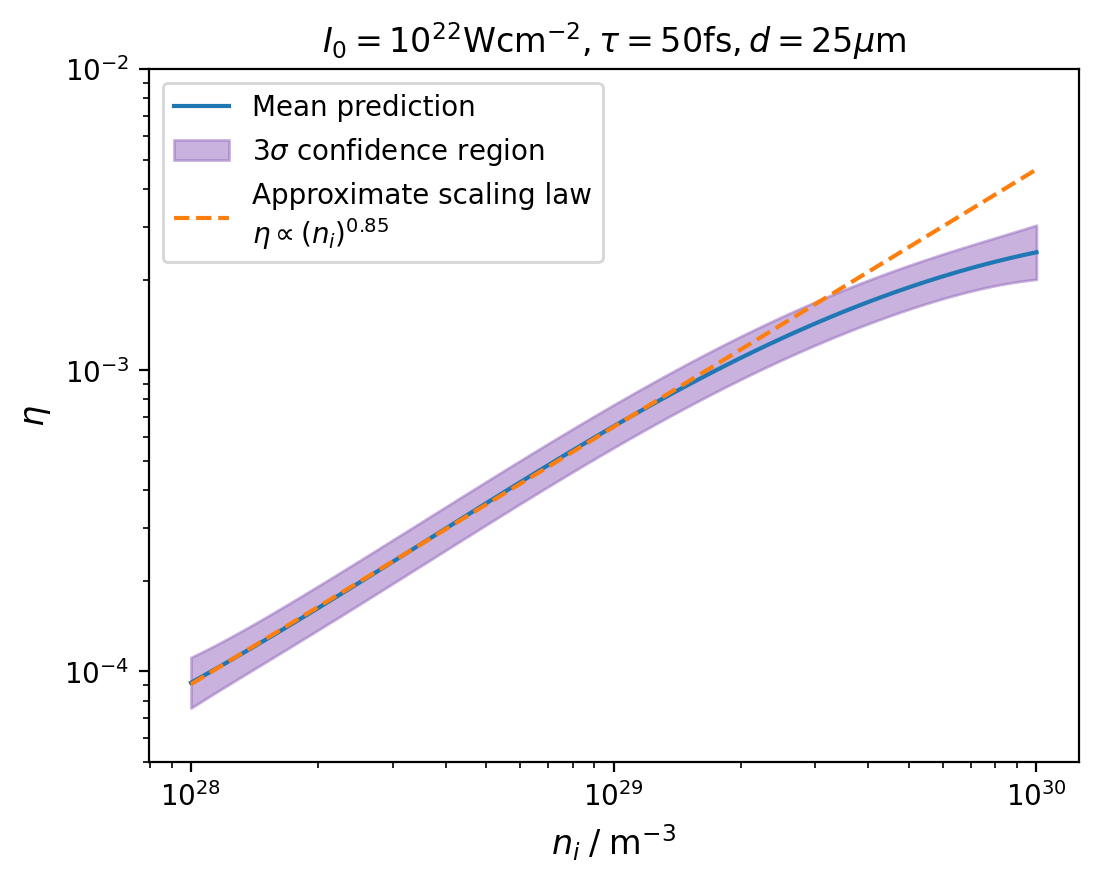}
        \caption{Target density}
    \end{subfigure}
    \caption{Example scans across the parameter space using the trained model, for each of the four variable parameters. Scaling laws are fitted to these scans to investigate how close the model is to the approximate analytical model. }
    \label{fig:parameter_scans}
\end{figure*}

The full model outlined at the start of this section has been used to produce predictions of conversion efficiency for full scans of each of the four parameters. These are shown in Figure (\ref{fig:parameter_scans}), where each parameter is scanned over while the other three parameters in each instance are kept fixed. Without accounting for any other effects, the maximum possible conversion efficiency in the simulation set up is $\eta=0.3$, i.e. the laser-to-electron efficiency, corresponding to all electrons emitting the entirety of their energy as bremsstrahlung. This will not be reached however due to multiple other effects. The nature of the TNSA boundaries used will mean approximately $22\%$ of electrons will escape the simulation quickly, and some energy will be lost by all electrons upon reflection by the TNSA boundaries. Additionally, other energy loss mechanisms are dominant for low energies and so we expect much lower conversion efficiencies, especially for a small $Z$ value. 

The variation of conversion efficiency with the different parameters is quantified by fitting scaling laws to the mean prediction across a scan of the parameter of interest. This can only be done for the low-emission end of the scale, as for three of the four parameters the conversion efficiency begins to tend to a maximum as the parameter increases. The calculated fits are shown alongside the sample scans in Figure \ref{fig:parameter_scans}. 

The behaviour of model with changing pulse length matches expectations; a longer pulse length just increases the number of electrons and total energy, without changing the electron dynamics greatly. Because of this, the conversion efficiency is mostly flat. For very small pulse lengths, the total number of injected electron macroparticles is comparatively low due to the pulse length approaching a similar scale to the simulation time step. As such statistical noise will increase, resulting in the high error in this region.

\subsection{\label{subsec:resolution}Variation with simulation hyperparameters}
The model above is trained on simulations using 40nm wide cells. To test how differing simulation resolutions change the results, the simulations were repeated at the same points in parameter space with grid cell sizes of 100nm. The conversion efficiencies calculated for this are shown in Figure (\ref{fig:summary_2}). These are, generally, very similar to the efficiencies calculated for a smaller cell size. When a GPR model is fitted to this data, the kernel hyperparameters are found to have an optimum of $a = 1.260, l = 2.900$ and $w=2.23\times10^{-3}$. From this we can see a relatively small change in the main shape parameters, with a more significant change in the noise parameter, with it increasing by a factor of approximately 2.34. The overall trend seen in the previous dataset as $\alpha$ is varied, that being that the white noise parameter and alpha sum to a constant value, is also seen here but with an appropriately larger ``true noise", as shown in Figure (\ref{fig:alpha}). Convergence of the model however occurs at approximately the same number of total simulations, with differing parameters following convergence, as shown in Figure (\ref{fig:param_convergence}). This seems to indicate that despite the increased noise, a consistent model can be still be reached quickly with lower resolution. Finally, as can be seen in Figure (\ref{Accuracy}), model accuracy experiences the same increase after approximately 10 simulations, however following this the calculated error in the model decreases much slower than in the higher resolution simulations.

\section{\label{sec:discuss}Discussion}
In the study of laser-plasma interactions, PIC codes are a ubiquitous tool for modelling the physics involved, with use in many different regimes and setups. However, they do not come without issues, namely the long simulation times required. This expensive computational aspect makes parameter scans impractical; A grid search of a four-dimensional parameter space, with 10 values across each dimension, results in 10,000 data points to simulate. This would be impractical, even in the case of 1D simulations. One other major issue is the noise of simulations. The number of simulated particles is vastly smaller than in a real interaction, and so due to the random nature of emission and other processes the electrons undergo, there will be some amount of noise in the results. This variation means that often simulations will need to be repeated multiple times in order to check if the simulation results aren't just random noise, further increasing the computation time required. 

As shown here however, it is possible to reduce the required data by a huge margin. This is due to the bayesian nature of the GPR model; an appropriate choice of prior vastly simplifies the model fitting process, and the model is optimized during this. Figure (\ref{fig:param_convergence}) show's that for the setup used in this work, this optimization process converges after approximately 400 data points. Additionally, accurate predictions can be made even early than this, as the models have high accuracy after very little data, as can be seen in Figure (\ref{Accuracy}). Furthermore, this is through a crude method; the data points are randomly sampled from the parameter space via a log-uniform distribution. This can be greatly improved upon within an ``active learning" scheme, where the next point to be sampled is chosen by finding the maximum value of an acquisition function. In our case where we are interested in exploring a parameter space, this function would simply be the model noise, such that uncertainty is minimized. In cases where optimization of a system is desired, the mean prediction can be used alongside the noise, with the latter weighted depending on how strongly exploration is wished for compared to exploitation.

The time savings of using a model such as this are significant; both datasets required thousands of CPU-hours to compute, while model training took on the order of a minute and prediction a fraction of a second. Because of this extreme time saving, models such as this are useful in cases where prediction is desired for a very large number of data points. One could potentially increase cell size further in order to reduce simulation time, however this can only be done so much before cell sizes begin to approach the size of the simulation space, or the time step approaches the timespan of electron injection. In both cases this will result in a poor approximation of the underlying system, and will still require significantly more computation time than our model, as even with a cell size of 1 micron, an average simulation in our dataset would take 10 CPU-minutes while poorly resolving the system.

The advantages of GPR also allow easy accounting for the noise of the simulations. As in our demonstration, the noise of the simulations can be directly quantified by including it within the GPR kernel. This additionally allows us to further refine our exploration method by subtracting the noise from the white kernel from the noise on a prediction, allowing a separation between the statistical noise of the simulation and the epistemic uncertainty resulting from a lack of sampling in a region. One matter to address with this however is that the current implementation considers the noise constant across parameter space. Already we can see this is not entirely accurate, as for small pulse lengths the uncertainty in results increases dramatically. This is likely due to the pulse length beginning to approach the value of the timestep, and as such the injection of electrons is poorly resolved. Variation of noise across the parameter space such as this could be accounted for in the model by using a white noise kernel that varies, preferably with this variation determined by hyperparameters that can be optimized.

Although the system looked at in this paper is relatively simple, it is a step towards a more complex framework for creating surrogate models from PIC data. Already it accomplishes the three requirements outlined in the introduction, those being quick evaluation, accurate interpolation and robustness to noise. It is a relatively simple matter to implement additional variable parameters in future models, especially in the case of using 2D PIC, where additional variables such as target width, spot size and incident angle become available. In these cases, the need for an active learning approach becomes necessary as computation times are increased even further in 2D and beyond. The computational efficiency of such methods can be improved further by using a more sophisticated method of choosing the initial data points than uniform random sampling, such as Latin Hypercube Sampling \cite{10.1214/aos/1069362310}.

Some other major limitations of the method are also imposed as consequences of the simulations used. Some parts of the simulation may not match the behaviour of real physics, whether this be due to approximation or for a lack of empirical data on the physics in that region. In the former case we have the implementation of the TNSA boundaries, where a specific escape energy and energy loss is defined for each simulation as being proportional to the mean hot electron energy. The proportionality constant of this relation is the same in all simulations, however in previous work such as \cite{10.1063/5.0055398}, this has been found to not be the case. Not including this variation changes the collective dynamics of the hot electrons and so will affect the bremsstrahlung result. One example of the latter issue is the physics of electron injection at the high end of our intensity range, where it hasn't been shown that the hot electron temperature follows the scaling used in our simulations and as such phenomena such as relativistic transparency could change this behaviour. Once again, this will change the overall dynamics of the electrons and so our results at high intensities may not match real behaviour. 

Finally, this method shows promise in improving some of the limitations of PIC codes, in particular the treatment of boundary behaviour in hybrid-PIC. While the escape energy and energy loss are parameterized as proportional to the mean energy of the injected electrons, the proportionality constant calculated from traditional PIC is dependent on the system parameters. The exact relationship however isn't well quantified, and often, for instance in this investigation, the values of $\kappa_{esc}$ and $\kappa_{TNSA}$ are chosen somewhat arbitrarily. GPR can be used as a tool to better choose such values in the future, by using the methods outlined above to produce a surrogate model for the two values across the parameter space. This can then be used alongside the hybrid-PIC code in the future to better model the physics of electrons at the boundary.

\section{\label{sec:conc}Conclusion}
In this paper, an investigation in to the use of Gaussian Process Regression for producing surrogate models of laser-plasma interactions has been performed. As a simple example of this, simulations were performed in 1D hybrid-PIC of the emission of bremsstrahlung by hot electrons in a plastic target. The data from these simulations was used to train a surrogate model using Gaussian Process Regression, using a square exponential kernel with added white noise. After fitting using the training data, this was then evaluated on how well it matches an analytical approximation of the bremsstrahlung emission. Other measures of the efficacy of the method were looked at, such as statistical noise and convergence of the kernel hyperparameters during training. Further improvements that will be made to the model are discussed, namely that of implementing active learning via an acquisition function, as well as potential use cases, such as building a surrogate model for the process of Target Normal Sheath Acceleration. 

\nocite{*}

\bibliography{apssamp}

\providecommand{\noopsort}[1]{}\providecommand{\singleletter}[1]{#1}%
\begin{thebibliography}{20}%
\makeatletter
\providecommand \@ifxundefined [1]{%
 \@ifx{#1\undefined}
}%
\providecommand \@ifnum [1]{%
 \ifnum #1\expandafter \@firstoftwo
 \else \expandafter \@secondoftwo
 \fi
}%
\providecommand \@ifx [1]{%
 \ifx #1\expandafter \@firstoftwo
 \else \expandafter \@secondoftwo
 \fi
}%
\providecommand \natexlab [1]{#1}%
\providecommand \enquote  [1]{``#1''}%
\providecommand \bibnamefont  [1]{#1}%
\providecommand \bibfnamefont [1]{#1}%
\providecommand \citenamefont [1]{#1}%
\providecommand \href@noop [0]{\@secondoftwo}%
\providecommand \href [0]{\begingroup \@sanitize@url \@href}%
\providecommand \@href[1]{\@@startlink{#1}\@@href}%
\providecommand \@@href[1]{\endgroup#1\@@endlink}%
\providecommand \@sanitize@url [0]{\catcode `\\12\catcode `\$12\catcode `\&12\catcode `\#12\catcode `\^12\catcode `\_12\catcode `\%12\relax}%
\providecommand \@@startlink[1]{}%
\providecommand \@@endlink[0]{}%
\providecommand \url  [0]{\begingroup\@sanitize@url \@url }%
\providecommand \@url [1]{\endgroup\@href {#1}{\urlprefix }}%
\providecommand \urlprefix  [0]{URL }%
\providecommand \Eprint [0]{\href }%
\providecommand \doibase [0]{https://doi.org/}%
\providecommand \selectlanguage [0]{\@gobble}%
\providecommand \bibinfo  [0]{\@secondoftwo}%
\providecommand \bibfield  [0]{\@secondoftwo}%
\providecommand \translation [1]{[#1]}%
\providecommand \BibitemOpen [0]{}%
\providecommand \bibitemStop [0]{}%
\providecommand \bibitemNoStop [0]{.\EOS\space}%
\providecommand \EOS [0]{\spacefactor3000\relax}%
\providecommand \BibitemShut  [1]{\csname bibitem#1\endcsname}%
\let\auto@bib@innerbib\@empty
\bibitem [{\citenamefont {Gales}\ and\ \citenamefont {team}(2015)}]{gales2015laser}%
  \BibitemOpen
  \bibfield  {author} {\bibinfo {author} {\bibfnamefont {S.}~\bibnamefont {Gales}}\ and\ \bibinfo {author} {\bibfnamefont {E.-N.}\ \bibnamefont {team}},\ }\bibfield  {title} {\bibinfo {title} {Laser driven nuclear science and applications: The need of high efficiency, high power and high repetition rate laser beams},\ }\href@noop {} {\bibfield  {journal} {\bibinfo  {journal} {The European Physical Journal Special Topics}\ }\textbf {\bibinfo {volume} {224}},\ \bibinfo {pages} {2631} (\bibinfo {year} {2015})}\BibitemShut {NoStop}%
\bibitem [{\citenamefont {Furch}\ \emph {et~al.}(2022)\citenamefont {Furch}, \citenamefont {Witting}, \citenamefont {Osolodkov}, \citenamefont {Schell}, \citenamefont {Schulz},\ and\ \citenamefont {Vrakking}}]{Furch_2022}%
  \BibitemOpen
  \bibfield  {author} {\bibinfo {author} {\bibfnamefont {F.~J.}\ \bibnamefont {Furch}}, \bibinfo {author} {\bibfnamefont {T.}~\bibnamefont {Witting}}, \bibinfo {author} {\bibfnamefont {M.}~\bibnamefont {Osolodkov}}, \bibinfo {author} {\bibfnamefont {F.}~\bibnamefont {Schell}}, \bibinfo {author} {\bibfnamefont {C.~P.}\ \bibnamefont {Schulz}},\ and\ \bibinfo {author} {\bibfnamefont {M.~J.~J.}\ \bibnamefont {Vrakking}},\ }\bibfield  {title} {\bibinfo {title} {High power, high repetition rate laser-based sources for attosecond science},\ }\href {https://doi.org/10.1088/2515-7647/ac74fb} {\bibfield  {journal} {\bibinfo  {journal} {Journal of Physics: Photonics}\ }\textbf {\bibinfo {volume} {4}},\ \bibinfo {pages} {032001} (\bibinfo {year} {2022})}\BibitemShut {NoStop}%
\bibitem [{\citenamefont {Baumann}\ and\ \citenamefont {Pukhov}(2019)}]{baumann2019laser}%
  \BibitemOpen
  \bibfield  {author} {\bibinfo {author} {\bibfnamefont {C.}~\bibnamefont {Baumann}}\ and\ \bibinfo {author} {\bibfnamefont {A.}~\bibnamefont {Pukhov}},\ }\bibfield  {title} {\bibinfo {title} {Laser-solid interaction and its potential for probing radiative corrections in strong-field quantum electrodynamics},\ }\href@noop {} {\bibfield  {journal} {\bibinfo  {journal} {Plasma Physics and Controlled Fusion}\ }\textbf {\bibinfo {volume} {61}},\ \bibinfo {pages} {074010} (\bibinfo {year} {2019})}\BibitemShut {NoStop}%
\bibitem [{\citenamefont {Fedeli}\ \emph {et~al.}(2021)\citenamefont {Fedeli}, \citenamefont {Sainte-Marie}, \citenamefont {Zaim}, \citenamefont {Th\'evenet}, \citenamefont {Vay}, \citenamefont {Myers}, \citenamefont {Qu\'er\'e},\ and\ \citenamefont {Vincenti}}]{PhysRevLett.127.114801}%
  \BibitemOpen
  \bibfield  {author} {\bibinfo {author} {\bibfnamefont {L.}~\bibnamefont {Fedeli}}, \bibinfo {author} {\bibfnamefont {A.}~\bibnamefont {Sainte-Marie}}, \bibinfo {author} {\bibfnamefont {N.}~\bibnamefont {Zaim}}, \bibinfo {author} {\bibfnamefont {M.}~\bibnamefont {Th\'evenet}}, \bibinfo {author} {\bibfnamefont {J.~L.}\ \bibnamefont {Vay}}, \bibinfo {author} {\bibfnamefont {A.}~\bibnamefont {Myers}}, \bibinfo {author} {\bibfnamefont {F.}~\bibnamefont {Qu\'er\'e}},\ and\ \bibinfo {author} {\bibfnamefont {H.}~\bibnamefont {Vincenti}},\ }\bibfield  {title} {\bibinfo {title} {Probing strong-field qed with doppler-boosted petawatt-class lasers},\ }\href {https://doi.org/10.1103/PhysRevLett.127.114801} {\bibfield  {journal} {\bibinfo  {journal} {Phys. Rev. Lett.}\ }\textbf {\bibinfo {volume} {127}},\ \bibinfo {pages} {114801} (\bibinfo {year} {2021})}\BibitemShut {NoStop}%
\bibitem [{\citenamefont {Macleod}(2022)}]{Macleod_2022}%
  \BibitemOpen
  \bibfield  {author} {\bibinfo {author} {\bibfnamefont {A.~J.}\ \bibnamefont {Macleod}},\ }\bibfield  {title} {\bibinfo {title} {From theory to precision modelling of strong-field qed in the transition regime},\ }\href {https://doi.org/10.1088/1742-6596/2249/1/012022} {\bibfield  {journal} {\bibinfo  {journal} {Journal of Physics: Conference Series}\ }\textbf {\bibinfo {volume} {2249}},\ \bibinfo {pages} {012022} (\bibinfo {year} {2022})}\BibitemShut {NoStop}%
\bibitem [{\citenamefont {Račiukaitis}(2021)}]{microfabrication}%
  \BibitemOpen
  \bibfield  {author} {\bibinfo {author} {\bibfnamefont {G.}~\bibnamefont {Račiukaitis}},\ }\bibfield  {title} {\bibinfo {title} {Ultra-short pulse lasers for microfabrication: A review},\ }\href {https://doi.org/10.1109/JSTQE.2021.3097009} {\bibfield  {journal} {\bibinfo  {journal} {IEEE Journal of Selected Topics in Quantum Electronics}\ }\textbf {\bibinfo {volume} {27}},\ \bibinfo {pages} {1} (\bibinfo {year} {2021})}\BibitemShut {NoStop}%
\bibitem [{\citenamefont {Kalaycıoğlu}\ \emph {et~al.}(2018)\citenamefont {Kalaycıoğlu}, \citenamefont {Elahi}, \citenamefont {Akçaalan},\ and\ \citenamefont {Ilday}}]{laser_ablation}%
  \BibitemOpen
  \bibfield  {author} {\bibinfo {author} {\bibfnamefont {H.}~\bibnamefont {Kalaycıoğlu}}, \bibinfo {author} {\bibfnamefont {P.}~\bibnamefont {Elahi}}, \bibinfo {author} {\bibfnamefont {Ã.}~\bibnamefont {Akçaalan}},\ and\ \bibinfo {author} {\bibfnamefont {F.~Ã.}\ \bibnamefont {Ilday}},\ }\bibfield  {title} {\bibinfo {title} {High-repetition-rate ultrafast fiber lasers for material processing},\ }\href {https://doi.org/10.1109/JSTQE.2017.2771745} {\bibfield  {journal} {\bibinfo  {journal} {IEEE Journal of Selected Topics in Quantum Electronics}\ }\textbf {\bibinfo {volume} {24}},\ \bibinfo {pages} {1} (\bibinfo {year} {2018})}\BibitemShut {NoStop}%
\bibitem [{\citenamefont {H{\"a}drich}\ \emph {et~al.}(2015)\citenamefont {H{\"a}drich}, \citenamefont {Krebs}, \citenamefont {Hoffmann}, \citenamefont {Klenke}, \citenamefont {Rothhardt}, \citenamefont {Limpert},\ and\ \citenamefont {T{\"u}nnermann}}]{hadrich2015exploring}%
  \BibitemOpen
  \bibfield  {author} {\bibinfo {author} {\bibfnamefont {S.}~\bibnamefont {H{\"a}drich}}, \bibinfo {author} {\bibfnamefont {M.}~\bibnamefont {Krebs}}, \bibinfo {author} {\bibfnamefont {A.}~\bibnamefont {Hoffmann}}, \bibinfo {author} {\bibfnamefont {A.}~\bibnamefont {Klenke}}, \bibinfo {author} {\bibfnamefont {J.}~\bibnamefont {Rothhardt}}, \bibinfo {author} {\bibfnamefont {J.}~\bibnamefont {Limpert}},\ and\ \bibinfo {author} {\bibfnamefont {A.}~\bibnamefont {T{\"u}nnermann}},\ }\bibfield  {title} {\bibinfo {title} {Exploring new avenues in high repetition rate table-top coherent extreme ultraviolet sources},\ }\href@noop {} {\bibfield  {journal} {\bibinfo  {journal} {Light: Science \& Applications}\ }\textbf {\bibinfo {volume} {4}},\ \bibinfo {pages} {e320} (\bibinfo {year} {2015})}\BibitemShut {NoStop}%
\bibitem [{\citenamefont {Alizadeh}\ \emph {et~al.}(2020)\citenamefont {Alizadeh}, \citenamefont {Allen},\ and\ \citenamefont {Mistree}}]{alizadeh2020managing}%
  \BibitemOpen
  \bibfield  {author} {\bibinfo {author} {\bibfnamefont {R.}~\bibnamefont {Alizadeh}}, \bibinfo {author} {\bibfnamefont {J.~K.}\ \bibnamefont {Allen}},\ and\ \bibinfo {author} {\bibfnamefont {F.}~\bibnamefont {Mistree}},\ }\bibfield  {title} {\bibinfo {title} {Managing computational complexity using surrogate models: a critical review},\ }\href@noop {} {\bibfield  {journal} {\bibinfo  {journal} {Research in Engineering Design}\ }\textbf {\bibinfo {volume} {31}},\ \bibinfo {pages} {275} (\bibinfo {year} {2020})}\BibitemShut {NoStop}%
\bibitem [{\citenamefont {Goodman}\ \emph {et~al.}(2023)\citenamefont {Goodman}, \citenamefont {King}, \citenamefont {Dolier}, \citenamefont {Wilson}, \citenamefont {Gray},\ and\ \citenamefont {McKenna}}]{Goodman_King_Dolier_Wilson_Gray_McKenna_2023}%
  \BibitemOpen
  \bibfield  {author} {\bibinfo {author} {\bibfnamefont {J.}~\bibnamefont {Goodman}}, \bibinfo {author} {\bibfnamefont {M.}~\bibnamefont {King}}, \bibinfo {author} {\bibfnamefont {E.~J.}\ \bibnamefont {Dolier}}, \bibinfo {author} {\bibfnamefont {R.}~\bibnamefont {Wilson}}, \bibinfo {author} {\bibfnamefont {R.~J.}\ \bibnamefont {Gray}},\ and\ \bibinfo {author} {\bibfnamefont {P.}~\bibnamefont {McKenna}},\ }\bibfield  {title} {\bibinfo {title} {Optimization and control of synchrotron emission in ultraintense laser–solid interactions using machine learning},\ }\href {https://doi.org/10.1017/hpl.2023.11} {\bibfield  {journal} {\bibinfo  {journal} {High Power Laser Science and Engineering}\ }\textbf {\bibinfo {volume} {11}},\ \bibinfo {pages} {e34} (\bibinfo {year} {2023})}\BibitemShut {NoStop}%
\bibitem [{\citenamefont {Goodman}\ \emph {et~al.}(2024)\citenamefont {Goodman}, \citenamefont {King}, \citenamefont {Dolier}, \citenamefont {Wilson}, \citenamefont {Gray},\ and\ \citenamefont {McKenna}}]{Goodman_King_Dolier_Wilson_Gray_McKenna_2024}%
  \BibitemOpen
  \bibfield  {author} {\bibinfo {author} {\bibfnamefont {J.}~\bibnamefont {Goodman}}, \bibinfo {author} {\bibfnamefont {M.}~\bibnamefont {King}}, \bibinfo {author} {\bibfnamefont {E.~J.}\ \bibnamefont {Dolier}}, \bibinfo {author} {\bibfnamefont {R.}~\bibnamefont {Wilson}}, \bibinfo {author} {\bibfnamefont {R.~J.}\ \bibnamefont {Gray}},\ and\ \bibinfo {author} {\bibfnamefont {P.}~\bibnamefont {McKenna}},\ }\bibfield  {title} {\bibinfo {title} {Optimization and control of synchrotron emission in ultraintense laser–solid interactions using machine learning – corrigendum},\ }\href {https://doi.org/10.1017/hpl.2023.95} {\bibfield  {journal} {\bibinfo  {journal} {High Power Laser Science and Engineering}\ }\textbf {\bibinfo {volume} {12}},\ \bibinfo {pages} {e12} (\bibinfo {year} {2024})}\BibitemShut {NoStop}%
\bibitem [{\citenamefont {Streeter}\ \emph {et~al.}(2023)\citenamefont {Streeter}, \citenamefont {Colgan}, \citenamefont {Cobo}, \citenamefont {Arran}, \citenamefont {Los}, \citenamefont {Watt}, \citenamefont {Bourgeois}, \citenamefont {Calvin}, \citenamefont {Carderelli}, \citenamefont {Cavanagh},\ and\ \citenamefont {et~al.}}]{Streeter_Colgan_Cobo_Arran_Los_Watt_Bourgeois_Calvin_Carderelli_Cavanagh_et_al._2023}%
  \BibitemOpen
  \bibfield  {author} {\bibinfo {author} {\bibfnamefont {M.~J.~V.}\ \bibnamefont {Streeter}}, \bibinfo {author} {\bibfnamefont {C.}~\bibnamefont {Colgan}}, \bibinfo {author} {\bibfnamefont {C.~C.}\ \bibnamefont {Cobo}}, \bibinfo {author} {\bibfnamefont {C.}~\bibnamefont {Arran}}, \bibinfo {author} {\bibfnamefont {E.~E.}\ \bibnamefont {Los}}, \bibinfo {author} {\bibfnamefont {R.}~\bibnamefont {Watt}}, \bibinfo {author} {\bibfnamefont {N.}~\bibnamefont {Bourgeois}}, \bibinfo {author} {\bibfnamefont {L.}~\bibnamefont {Calvin}}, \bibinfo {author} {\bibfnamefont {J.}~\bibnamefont {Carderelli}}, \bibinfo {author} {\bibfnamefont {N.}~\bibnamefont {Cavanagh}},\ and\ \bibinfo {author} {\bibnamefont {et~al.}},\ }\bibfield  {title} {\bibinfo {title} {Laser wakefield accelerator modelling with variational neural networks},\ }\href {https://doi.org/10.1017/hpl.2022.47} {\bibfield  {journal} {\bibinfo  {journal} {High Power Laser Science and Engineering}\ }\textbf {\bibinfo {volume} {11}},\ \bibinfo {pages} {e9} (\bibinfo
  {year} {2023})}\BibitemShut {NoStop}%
\bibitem [{\citenamefont {Watt}\ \emph {et~al.}(2023)\citenamefont {Watt}, \citenamefont {Rose}, \citenamefont {Kettle},\ and\ \citenamefont {Mangles}}]{PhysRevAccelBeams.26.054601}%
  \BibitemOpen
  \bibfield  {author} {\bibinfo {author} {\bibfnamefont {R.~A.}\ \bibnamefont {Watt}}, \bibinfo {author} {\bibfnamefont {S.~J.}\ \bibnamefont {Rose}}, \bibinfo {author} {\bibfnamefont {B.}~\bibnamefont {Kettle}},\ and\ \bibinfo {author} {\bibfnamefont {S.~P.~D.}\ \bibnamefont {Mangles}},\ }\bibfield  {title} {\bibinfo {title} {Monte carlo modeling of the linear breit-wheeler process within the geant4 framework},\ }\href {https://doi.org/10.1103/PhysRevAccelBeams.26.054601} {\bibfield  {journal} {\bibinfo  {journal} {Phys. Rev. Accel. Beams}\ }\textbf {\bibinfo {volume} {26}},\ \bibinfo {pages} {054601} (\bibinfo {year} {2023})}\BibitemShut {NoStop}%
\bibitem [{\citenamefont {Jackson}(1999)}]{JacksonJohnDavid1925-20161999Ce/J}%
  \BibitemOpen
  \bibfield  {author} {\bibinfo {author} {\bibfnamefont {.-.}\ \bibnamefont {Jackson}, \bibfnamefont {John~David}},\ }\href@noop {} {\emph {\bibinfo {title} {Classical electrodynamics / John David Jackson.}}},\ \bibinfo {edition} {3rd}\ ed.\ (\bibinfo  {publisher} {Wiley},\ \bibinfo {address} {New York},\ \bibinfo {year} {1999})\BibitemShut {NoStop}%
\bibitem [{\citenamefont {Nilson}\ \emph {et~al.}(2010)\citenamefont {Nilson}, \citenamefont {Solodov}, \citenamefont {Myatt}, \citenamefont {Theobald}, \citenamefont {Jaanimagi}, \citenamefont {Gao}, \citenamefont {Stoeckl}, \citenamefont {Craxton}, \citenamefont {Delettrez}, \citenamefont {Yaakobi}, \citenamefont {Zuegel}, \citenamefont {Kruschwitz}, \citenamefont {Dorrer}, \citenamefont {Kelly}, \citenamefont {Akli}, \citenamefont {Patel}, \citenamefont {Mackinnon}, \citenamefont {Betti}, \citenamefont {Sangster},\ and\ \citenamefont {Meyerhofer}}]{PhysRevLett.105.235001}%
  \BibitemOpen
  \bibfield  {author} {\bibinfo {author} {\bibfnamefont {P.~M.}\ \bibnamefont {Nilson}}, \bibinfo {author} {\bibfnamefont {A.~A.}\ \bibnamefont {Solodov}}, \bibinfo {author} {\bibfnamefont {J.~F.}\ \bibnamefont {Myatt}}, \bibinfo {author} {\bibfnamefont {W.}~\bibnamefont {Theobald}}, \bibinfo {author} {\bibfnamefont {P.~A.}\ \bibnamefont {Jaanimagi}}, \bibinfo {author} {\bibfnamefont {L.}~\bibnamefont {Gao}}, \bibinfo {author} {\bibfnamefont {C.}~\bibnamefont {Stoeckl}}, \bibinfo {author} {\bibfnamefont {R.~S.}\ \bibnamefont {Craxton}}, \bibinfo {author} {\bibfnamefont {J.~A.}\ \bibnamefont {Delettrez}}, \bibinfo {author} {\bibfnamefont {B.}~\bibnamefont {Yaakobi}}, \bibinfo {author} {\bibfnamefont {J.~D.}\ \bibnamefont {Zuegel}}, \bibinfo {author} {\bibfnamefont {B.~E.}\ \bibnamefont {Kruschwitz}}, \bibinfo {author} {\bibfnamefont {C.}~\bibnamefont {Dorrer}}, \bibinfo {author} {\bibfnamefont {J.~H.}\ \bibnamefont {Kelly}}, \bibinfo {author} {\bibfnamefont {K.~U.}\ \bibnamefont {Akli}}, \bibinfo {author}
  {\bibfnamefont {P.~K.}\ \bibnamefont {Patel}}, \bibinfo {author} {\bibfnamefont {A.~J.}\ \bibnamefont {Mackinnon}}, \bibinfo {author} {\bibfnamefont {R.}~\bibnamefont {Betti}}, \bibinfo {author} {\bibfnamefont {T.~C.}\ \bibnamefont {Sangster}},\ and\ \bibinfo {author} {\bibfnamefont {D.~D.}\ \bibnamefont {Meyerhofer}},\ }\bibfield  {title} {\bibinfo {title} {Scaling hot-electron generation to high-power, kilojoule-class laser-solid interactions},\ }\href {https://doi.org/10.1103/PhysRevLett.105.235001} {\bibfield  {journal} {\bibinfo  {journal} {Phys. Rev. Lett.}\ }\textbf {\bibinfo {volume} {105}},\ \bibinfo {pages} {235001} (\bibinfo {year} {2010})}\BibitemShut {NoStop}%
\bibitem [{\citenamefont {Morris}\ \emph {et~al.}(2021)\citenamefont {Morris}, \citenamefont {Robinson},\ and\ \citenamefont {Ridgers}}]{10.1063/5.0055398}%
  \BibitemOpen
  \bibfield  {author} {\bibinfo {author} {\bibfnamefont {S.}~\bibnamefont {Morris}}, \bibinfo {author} {\bibfnamefont {A.}~\bibnamefont {Robinson}},\ and\ \bibinfo {author} {\bibfnamefont {C.}~\bibnamefont {Ridgers}},\ }\bibfield  {title} {\bibinfo {title} {{Highly efficient conversion of laser energy to hard x-rays in high-intensity laser–solid simulations}},\ }\href {https://doi.org/10.1063/5.0055398} {\bibfield  {journal} {\bibinfo  {journal} {Physics of Plasmas}\ }\textbf {\bibinfo {volume} {28}},\ \bibinfo {pages} {103304} (\bibinfo {year} {2021})},\ \Eprint {https://arxiv.org/abs/https://pubs.aip.org/aip/pop/article-pdf/doi/10.1063/5.0055398/19332218/103304\_1\_5.0055398.pdf} {https://pubs.aip.org/aip/pop/article-pdf/doi/10.1063/5.0055398/19332218/103304\_1\_5.0055398.pdf} \BibitemShut {NoStop}%
\bibitem [{\citenamefont {Rusby}\ \emph {et~al.}(2019)\citenamefont {Rusby}, \citenamefont {Armstrong}, \citenamefont {Scott}, \citenamefont {King}, \citenamefont {McKenna},\ and\ \citenamefont {Neely}}]{Rusby_Armstrong_Scott_King_McKenna_Neely_2019}%
  \BibitemOpen
  \bibfield  {author} {\bibinfo {author} {\bibfnamefont {D.~R.}\ \bibnamefont {Rusby}}, \bibinfo {author} {\bibfnamefont {C.~D.}\ \bibnamefont {Armstrong}}, \bibinfo {author} {\bibfnamefont {G.~G.}\ \bibnamefont {Scott}}, \bibinfo {author} {\bibfnamefont {M.}~\bibnamefont {King}}, \bibinfo {author} {\bibfnamefont {P.}~\bibnamefont {McKenna}},\ and\ \bibinfo {author} {\bibfnamefont {D.}~\bibnamefont {Neely}},\ }\bibfield  {title} {\bibinfo {title} {Effect of rear surface fields on hot, refluxing and escaping electron populations via numerical simulations},\ }\href {https://doi.org/10.1017/hpl.2019.34} {\bibfield  {journal} {\bibinfo  {journal} {High Power Laser Science and Engineering}\ }\textbf {\bibinfo {volume} {7}},\ \bibinfo {pages} {e45} (\bibinfo {year} {2019})}\BibitemShut {NoStop}%
\bibitem [{\citenamefont {Rasmussen}\ and\ \citenamefont {Williams}(2006)}]{RasmussenCarlEdward2006Gpfm}%
  \BibitemOpen
  \bibfield  {author} {\bibinfo {author} {\bibfnamefont {C.~E.}\ \bibnamefont {Rasmussen}}\ and\ \bibinfo {author} {\bibfnamefont {C.~K.~I.}\ \bibnamefont {Williams}},\ }\href@noop {} {\emph {\bibinfo {title} {Gaussian processes for machine learning / Carl Edward Rasmussen, Christopher K.I. Williams.}}},\ Adaptive computation and machine learning\ (\bibinfo  {publisher} {MIT Press},\ \bibinfo {address} {Cambridge, Mass.},\ \bibinfo {year} {2006})\BibitemShut {NoStop}%
\bibitem [{\citenamefont {Loh}(1996)}]{10.1214/aos/1069362310}%
  \BibitemOpen
  \bibfield  {author} {\bibinfo {author} {\bibfnamefont {W.-L.}\ \bibnamefont {Loh}},\ }\bibfield  {title} {\bibinfo {title} {{On Latin hypercube sampling}},\ }\href {https://doi.org/10.1214/aos/1069362310} {\bibfield  {journal} {\bibinfo  {journal} {The Annals of Statistics}\ }\textbf {\bibinfo {volume} {24}},\ \bibinfo {pages} {2058 } (\bibinfo {year} {1996})}\BibitemShut {NoStop}%
\bibitem [{\citenamefont {Arber}\ \emph {et~al.}(2015)\citenamefont {Arber}, \citenamefont {Bennett}, \citenamefont {Brady}, \citenamefont {Lawrence-Douglas}, \citenamefont {Ramsay}, \citenamefont {Sircombe}, \citenamefont {Gillies}, \citenamefont {Evans}, \citenamefont {Schmitz}, \citenamefont {Bell},\ and\ \citenamefont {Ridgers}}]{Arber_2015}%
  \BibitemOpen
  \bibfield  {author} {\bibinfo {author} {\bibfnamefont {T.~D.}\ \bibnamefont {Arber}}, \bibinfo {author} {\bibfnamefont {K.}~\bibnamefont {Bennett}}, \bibinfo {author} {\bibfnamefont {C.~S.}\ \bibnamefont {Brady}}, \bibinfo {author} {\bibfnamefont {A.}~\bibnamefont {Lawrence-Douglas}}, \bibinfo {author} {\bibfnamefont {M.~G.}\ \bibnamefont {Ramsay}}, \bibinfo {author} {\bibfnamefont {N.~J.}\ \bibnamefont {Sircombe}}, \bibinfo {author} {\bibfnamefont {P.}~\bibnamefont {Gillies}}, \bibinfo {author} {\bibfnamefont {R.~G.}\ \bibnamefont {Evans}}, \bibinfo {author} {\bibfnamefont {H.}~\bibnamefont {Schmitz}}, \bibinfo {author} {\bibfnamefont {A.~R.}\ \bibnamefont {Bell}},\ and\ \bibinfo {author} {\bibfnamefont {C.~P.}\ \bibnamefont {Ridgers}},\ }\bibfield  {title} {\bibinfo {title} {Contemporary particle-in-cell approach to laser-plasma modelling},\ }\href {https://doi.org/10.1088/0741-3335/57/11/113001} {\bibfield  {journal} {\bibinfo  {journal} {Plasma Physics and Controlled Fusion}\ }\textbf {\bibinfo {volume}
  {57}},\ \bibinfo {pages} {113001} (\bibinfo {year} {2015})}\BibitemShut {NoStop}%
\end{thebibliography}%

\end{document}